# Blending of Light in Gravitational Microlensing Events

R. Di Stefano and A. A. Esin

Harvard-Smithsonian Center for Astrophysics

60 Garden st, Cambridge, MA 02138

**Abstract**

When there is more than one source of light along the line of sight to a gravitationally lensed object, the characteristics of the observed light curve are influenced by the presence of the light that is not lensed. In this paper we develop a formalism to quantify the associated effects. We find it useful to introduce the concept of a "blended Einstein radius" and an "effective Einstein radius", to describe the probability that a mass will serve as a lens, or that a source will be lensed in an observable way.

These considerations lead to generic predictions about the results of gravitational microlensing experiments. One example is that the optical depth for the lensing of giants is greater than that for the lensing of main sequence stars; for any given population of sources and lenses this effect can be quantified. We test and sharpen these predictions by performing a series of Monte Carlo simulations. We also outline general methods to (1) test whether specific events which fail to be fit by point-mass light curves are viable candidates for blended events, (2) use the effects of blending to learn more about the lensing event than would be possible if there were no blending, and (3) include the effects of blending when inferring properties of underlying populations through the statistical study of lensing events.

## 1 Introduction

Observational programs to detect microlensing events monitor $\sim 10^6 - 10^7$ stars per night (see, e.g., Alcock et al. 1993; Aubourg et al. 1993; Udalski et al. 1993). It is likely that in a significant fraction of these cases, the light received by our detectors is blended. The lensed source could be in a binary, for example, or there could be other stars within the resolution cone of the telescope that are not physically associated with the lensed source. Whatever the origin of the additional light, its presence has implications for what is observed (both in individual events and in the ensemble of events), and how the observations are interpreted. In §2 we outline the general effects of blending and develop a useful formalism to estimate the significance of the role of blending in observational programs designed to detect microlensing. Our estimates are tested in a set of Monte Carlo simulations. Section 3 is devoted to a brief discussion of our results and their implications.

## 2 Quantifying the Effects of Blending

If both the lens and lensed source can be approximated as point masses, then the amplification, $A$, can be expressed in terms of the distance, $u$, from the lens to the projection of the source in the lens plane. It is convenient to express $u$ in units of the Einstein radius:

$R_E = \sqrt{[4G\,m\,D\,x\,(1-x)/c^2]}$, where $D$ denotes the distance of the lensed source from the observer, and $xD$ denotes the distance of the lens. We have $A(u) = (u^2 + 2)/(u\sqrt{u^2 + 4})$. When $u \leq 1$, $A(u) \geq e$, where $e \sim 1.34$. We will refer to $A(u)$ as the "true" amplification; it is independent of the wavelength of the lensed light. When there are other sources of light along the same line of sight, then the light from the lensed source constitutes only a fraction, $f$, of the baseline luminosity, $L$. During the lensing event, only $fL$ is amplified. The relationship between the observed amplification, $A^{obs}$ and the true amplification is

$$(A^{obs}(u) - 1) = (A(u) - 1)\,f \tag{1}$$

Since the spectrum of the lensed light is generally different from that of the unlensed light, $f$ and $A^{obs}$ will be wavelength dependent. In addition to possible chromatic effects, Eq. (1) implies that the effects of blending cause an event to mimic another with a smaller Einstein radius and a larger distance of closest approach.

## 2.1 The "Blended" Einstein Radius

Unlike physical quantities such as the mass of a particle or its charge, which are generally viewed as intrinsic properties, the Einstein radius is a function not only of something characteristic of the lens, its mass, but also of variables related to its environment, specifically, its relative placement with respect to the observer and the source of deflected light. In this sense it has something in common with physical quantities like the charge associated with an electron as it moves through matter, whose effect is influenced by the configuration and properties of the material the electron travels through.

When there is blending, it is useful to define a quantity $R_{E,b}$, the blended Einstein radius. When $u \leq \frac{R_{E,b}}{R_E}$, then $A^{obs} \geq e = 1.34$. An expression for $R_{E,b}$ is readily derived:

$$R_{E,b} = R_E\sqrt{2}\sqrt{\frac{(e-1)+f}{\sqrt{(e-1)^2 + 2(e-1)\,f}} - 1} \tag{2}$$

Figure 1 shows a plot of $R_{E,b}$ and $R_{E,b}^2$, as functions of $f$.

Figure 1. The blended Einstein radius (in units of $R_E$) and its square (in units of $R_E^2$), as a function of the fraction of light contributed by the lensed source. (See Eq. (2).)

The concept of a "blended" Einstein radius augments both the intrinsic lens properties and the environmental factors included in $R_E$. Its value is influenced by the intrinsic luminosity of all sources along the line of sight, as well the relative distances.

## 2.2 The Effective Einstein Radius

Figure (1) illustrates that for any source-lens pair, the chance of observing an event is decreased if there is blending, since $R_{E,b} < R_E$ if $f < 1$. The overall effect of blending, however, cannot be estimated simply from Eq. (2). This is because blending causes some sources that would, by themselves be too dim to be monitored, to be included in the observations, precisely because there *are* other sources of light along the same line of sight. These less luminous sources are typically more numerous. Hence, in order to develop a complete picture of the effects of blending, we must include population effects. It therefore seems appropriate to expand upon the analogy between the Einstein radius and the charge of an electron as it moves through matter. In the latter case, the role played by the environment is taken into account by defining an effective charge. For gravitational lenses, and we define an *effective* Einstein radius. To do this, we take a population-weighted average over some properties, to derive $R_{E,b}^{eff}$ as a function of other properties (such as mass) that might be of interest to a particular investigation. The concept is illustrated below for the case of lensing by luminous stars; it can be equally well applied to other situations, such as blending due to lensing of binaries (which has been studied in detail by Griest & Hu [1992]), and blending due to resolution limits, which we will comment on in §3.

Let $\rho(m_l)$ [$\rho(m_s)$] represent the normalized number density associated with the IMF of the lens [source] population. In order to treat the relative distances simply, we model the spatial distribution of sources along the line of sight as a $\delta$-function at some fixed distance, $D$, and that of the lenses as a $\delta$-function at a distance $xD$. We assume that the luminosity of each star is a function of its mass, $m$, and core mass, $m_c$: $L = L(m, m_c)$. If we limit ourselves to lines of sight along which the total flux is great enough that the total apparent magnitude $M_V^{total}$ is less than some $M_V^{max}$, then the effective Einstein radius as a function of lens mass is $R_{E,b}^{eff}(m_l) = \int dm_{c,l} \left[ \int dm_s \int dm_{c,s} \, \rho(m_s) \, R_{E,b}(m_l, m_s, L_l, L_s) \, \theta(M_V^{max} - M_V^{total}) \right]$. This function is plotted in panel (a) of Figure 2 for a parructular example (our "standard model") described in the caption[1]. Also shown (dashed line) is the effective Einstein radius with no blending, $R_E^{eff}(m_l) = \int dm_s \rho(m_s) \, R_E(m_l) \, \theta(M_V^{max} - M_V^{source})$. In contrast to $R_E^{eff}(m_l)$, $R_{E,b}^{eff}(m_l)$ has a well-defined maximum and the contribution of higher mass lenses is suppressed. The excess of $R_{E,b}^{eff}(m_l)$ near its maximum relative to $R_E^{eff}(m_l)$ is due to the lensing of the sources that lie just above the magnitude limit. To compute the rate[2] of lensing due to stars, we integrate $\int dm_l \, \rho(m_l) \, R_{E,b}^{eff}(m_l)$. The rate obtained in this way is $\sim 82\%$ of the rate that would be derived for the same lens and source populations if blending were not taken into account; i.e., it is $\sim \left[ 0.82 \int dm_l \, \rho(m_l) \, R_E^{eff}(m_l) \right]$. Decreasing $M_V^{max}$ by two magnitudes, makes a small change ($\sim 5\%$) in the result. Although, in this example, blending leads to a significant diminution of the rate of lensing due to stars, the result is sensitive to the details of the spatial and, especially of the mass distribution of lenses and sources; it is also possible for blending to lead to a net enhancement. The essential feature we derive from this simple example is a significant sensitivity to the form of the IMF in any realistic calculation of rates and optical depths.

---

[1] All numerical results refer to our standard model unless otherwise indicated.

[2] Note that formally, this calculation yields the *population-weighted-average* of the effective Einstein radius. If, in our simple $\delta$-function model of spatial distributions, we take the relative velocity to be constant, this quantity is a direct measure of the rate. Similarly, the optical depth is proportional to $\int dm_l \, \rho(m_l) \, [R_{E,b}^{eff}(m_l)]^2$.

Figure 2. Here the source and lens masses range from 0.1 to $3.0 M_\odot$, and are chosen from a Miller-Scalo IMF (Miller & Scalo 1979); The distances to the sources and lenses are fixed at 8 kpc and 4 kpc respectively. The magnitude cutoff is taken to be 23. (a) The effective blended Einstein radius as a function of lens mass is plotted as a solid line; the values obtained without blending, are plotted along the dotted line. Note that near its peak $R_{E,b}^{eff}(m_l)$ actually exceeds $R_E^{eff}(m_l)$. (b) The effective blended Einstein radius as a function of source mass $R_E^{eff}(m_s)$ is plotted as a solid line; the dashed line indicates the constant value that would have been derived through the same Monte Carlo integration without blending.

Note that the optical depth, $\tau$, can be written in terms of these effective Einstein radii[2]:
$\tau = \int dm_l [R_E^{eff}(m_l)]^2 = \int dm_s [R_E^{eff}(m_s)]^2$.

Whether or not the lens is luminous, blending is important when the field being monitored is crowded, as is always the case for observational programs designed to detect microlensing. If we assume that $N$ stars from the source population (whose combined luminosity places this line of sight below the magnitude limit), are typically within the resolution cone of the telescope, we find that the rate of lensing *increases* with $N$. This is because there are many possible lensing events associated with the low-mass stars in the resolution cone, instead of just one event associated with lensing of a single isolated star. On the other hand, the optical depth *decreases*, since each event is characterized by a blended Einstein radius that is smaller than $R_E$. For example, considering lensing by dark matter, with $\sim 25$ stars in the resolution cone (not a large number for the Galactic Bulge, assuming arc second resolution), we find that the rate is $\sim 50\%$ higher than if the resolution cone contained just a single visible star, while $\tau$ is $\sim 88\%$ of the value derived without blending. There is a marked difference in the duration of events as well, with many more events of short duration. We note that a complete picture of the effects of blending must include the results of computations of lensing by stars (both within the source population as well as between the source population and the Earth), as well as by dark matter, and also the effects of absorption and reddening.

An important characteristic of $R_{E,b}$ is that it depends on the properties of both the source and lens populations. One can therefore compute the effective Einstein radius as a function of *source* mass.
$R_{E,b}^{eff}(m_s) = \int dm_{c,s} \left[ \int dm_l \int dm_{c,l}\ \rho(m_l)\ R_{E,b}(m_l, m_s, L_l, L_s)\ \theta(M_V^{max} - M_V^{total}) \right]$. This function is plotted in panel (b) of Figure 2. It is clear that not all masses are equally likely to be lensed in an observable way, with lensing of more massive, hence more luminous stars, much more likely to be observed. Another interesting feature of this figure is the comparison with the $R_E^{eff}(m_s)$ which does not take blending into account (dashed line). Since, without blending, stars of the lowest mass will not be monitored, they will not contribute. With blending, we find that they contribute $\sim 12\%$ of all observed events. In fact, for all sources below $1 M_\odot$ (i.e., for $\sim 85\%$ of all the sources in our model), blending plays a significant role in making the lensing of their light either more or less observable.

## 2.3 Identifying Blending

It is important to identify events in which the incident light is blended, both to bolster statistical studies designed to learn about the population of lenses (see §2.3.2), and to use the blending to learn as much as possible about individual events. For example, if the presence of the additional light can be attributed to the lens itself, then a complete analysis of the event may allow us to determine the magnitude and spectral type of the lens, and thus its mass and distance from us. This allows us to compute the Einstein radius, so that the time duration of the event then establishes the value of the transverse velocity. Clearly, a collection of such events may give us valuable insight into the dynamics of the Galaxy. The extent to which this is possible for a single event has been explored by Udalski et al. (1994) and Alcock et al. (1995) for the OGLE 7 event (cf. Kamionkowski [1994]).

### 2.3.1 Methods for Individual Events

We discuss three approaches. The first depends only on the shape of the light curve, which is a function of $f$ and $b$ (the distance of closest approach in units of $R_E$), while the second and third depend on spectral measures.

We have used a modified version of the approach of Bolatto & Falco (1994) to identify the values of $f$ and $b$ for which blended events should be distinguishable through measurable deformation of the light curve relative to that of the unblended point-mass light curve that gives the best fit to the data. We find that $b$ is the parameter that has the most significant influence. The strong dependence on $b$ can be understood analytically, since one can show that for small $b$ it is not possible to find a fit that matches the function and its derivative both at the peak, and near $u = R_E$. Indeed it was such a mismatch that was responsible for the identification of the OGLE 7 binary lens event as a blended event (Udalski et al. 1994). A problem for future study of the light curves associated with point mass lenses is to identify the values of $b$ and $f$ (as a function of sampling frequency and photometry) for which unblended fits will not work. For now, we have used the general guideline that events for which $b < 0.15 - 0.2$ could possibly be identified as blended events because of anomalies in the light curve. Since, like the Einstein radius and its generalizations, this criterion is geometrical, it allows us to get a simple handle on the fraction of events for which blending may be detectable because of anomalies in the light curve. It is roughly $15 - 20\%$ of all events with $A^{true} > 1.34$, and is a larger fraction, $(15 - 20\%) \times \left(\frac{R_E}{\langle R_{E,b} \rangle}\right)$, of observable events.

The success of spectral measures requires not only a significant amount of blending, but also that there be a significant temperature difference between the lensed source and other sources of light along the line of sight. It is therefore more difficult to make predictions that are likely to be valid for all populations. We note however, that temperature differences are much more likely to be characteristic of events in which the source (lens) is relatively more massive and/or is relatively more evolved. Since more massive stars are less numerous and since advanced stages of evolution are passed through relatively quickly, blended events that are identified as such because of anomalies in spectral measures are likely to be only a small fraction of those detected because

Observing Signatures of Blended Events

| Event Type (S-L) Source-Lens | Percentage of All Observed Events With Type S-L[a] | Fraction of Observed Events of Type S-L Which Display a Characteristic Signature[b] | | | Percentage of Observed Events When Blending Is Discernible |
|---|---|---|---|---|---|
| | | $\|\Delta T\| > 500K$ | $\|\Delta(U-V)\| > 0.15$ | $b < 0.2$ | |
| MS-MS | 89.0% | 0.001 | 0.04 | 0.28 | 26.1% |
| MS-G | 0.5 | 0.078 | 0.91 | 0.58 | 0.4 |
| G-MS | 10.4 | 0.007 | 0.14 | 0.22 | 3.2 |
| G-G | 0.2 | 0.002 | 0.45 | 0.39 | 0.1 |
| | | | | | Total=29.8% |

Table 1: [a]These numbers are strongly dependent on the details of the model used. For example, we find that lowering the magnitude limit and using an IMF more appropriate to the Bulge dramatically increases the fraction of the G-MS events.
[b]Here we defined $\Delta T = T_{UV} - T_{BV}$.

of anomalies in their light curve. One measure of chromaticity depends on observed color changes during the event itself. The difference between the baseline color and the color at the maximum amplification is given by the expression $\Delta(C_1 - C_2) = -2.5 \log \left( \frac{f_{C_1}(A-1)+1}{f_{C_2}(A-1)+1} \right)$. Here $f_{C_1}$ and $f_{C_2}$ are the ratios of the flux from the lensed source to the total flux, in the color bands $C_1$ and $C_2$, respectively. The second spectroscopic method relies on studies of the baseline magnitude prior to and/or after the event. If the source is not blended, then its temperature, $T_{C_1 C_2}$, as determined from $C_1 - C_2$ will agree with its temperature, $T_{C_3 C_4}$, as determined from $C_3 - C_4$; a more detailed spectral analysis should confirm the derived temperature. If, however, there is a second source of a different temperature that provides a significant fraction of the incident light, then it is possible that $T_{C_1 C_2}$ may have a value that is significantly different from $T_{C_3 C_4}$, and that a more detailed spectral analysis will reveal anomalies. Such an analysis has the distinct advantage that it can be done without the time pressure associated with studying an event in progress.

Since the three measures are largely independent (although the value of $\Delta(C_1 - C_2)$ is related both to $b$ and to the temperature differences between the lensed and unlensed sources), it may be possible that the combination provides stronger constraints than would be possible if we had to rely on just one of them. To study the detectability of these blending signatures we performed a series of Monte Carlo simulations. The results for our standard model are summarized in Table 1.

### 2.3.2 Statistical Studies

Given an ensemble of lensing events, with little or no information about whether any individual members might be blended, one can nevertheless apply statistical measures to determine the likelihood that a significant fraction of the observed events were blended. The characteristic of an ensemble of observed events that is most obviously altered by blending, is the distribution of event durations. This distribution is more sharply peaked toward shorter durations if there is blending. Since, for a given value of the transverse velocity, and without taking blending into account, the event duration is proportional to $R_E \sim \sqrt{m_l}$, the distribution of event durations gives a measure of the distribution of lens masses, which will be artificially skewed toward low values if blending is a common feature. For example, although the true population of lenses in our model has no members with $m_l < 0.1 M_\odot$, 10% of the observed events would be associated with masses less than $0.1 M_\odot$, if we did not know that they were blended events. Further, although $\sim 50\%$ of the true lensing events are associated with lenses that are more massive than $0.5 M_\odot$, only 15% of the observed events would be of long enough duration that, without taking blending into account, we would infer $m_l > 0.5 M_\odot$.

Without an *a priori* knowledge of the lens population, how might one discover evidence of blending? One sign of blending (which seems to be present in the data) is that the optical depth associated with lensing of giants, $\tau_g$, should be larger than that associated with lensing of dwarfs, $\tau_{ms}$. In our standard model (without other stars in the resolution cone being taken into account), we find $\tau_g/\tau_{ms} > 1$, with the exact value of the ratio dependent on the magnitude cut-off. For the limiting magnitude of 19.5, the ratio is $\sim 2.5$. If we take blending due to $N$ unresolved stars in the source population into account, the value of the ratio becomes significantly smaller, approaching unity from above as $N$ increases.

Another sign of blending that can be studied with a large enough ensemble of events is the apparent lensing of stars in the lens population. (The MACHO group has observed $\sim 3$ such events.) These stars, should actually have a small probability of being lensed, since $x$ is close to unity. When blending is taken into account, we find that apparent "lensing of lenses" is likely to be due to lensing of stars in the more distant "source" population, if $L_{source}/L_{lens}$, as measured on Earth, is small. This conjecture is testable; in our standard model, we find that events for which $f < 0.2$ will have significantly smaller peak amplification and time duration. Without taking other effects (such as source confusion) into account, we find that the average value of the peak amplification of such events is a factor of $\sim 2$ lower than for the total population of events and the average time duration is reduced by a factor of $\sim 3$.

## 3  Discussion

The formalism described in §2 provides a conceptual framework within which it is straightforward to intuitively understand the effects of blending. The approach is geometrical. While the standard Einstein radius, $R_E$, determines the rate of microlensing by lenses of a given mass, the blended Einstein radius, $R_{E,b} < R_E$ determines the rate of observable events. We find the effective Einstein radius, $R_{E,b}^{eff}$, which is a population weighted average of $R_{E,b}$, to be a useful quantity. The identification of blended events through light curve anomalies can also be understood in a primarily geometrical way.

Blending leads to selection effects that are not inherent in the physics of gravitational lensing, but which are unavoidable in astrophysical observations. If blending is not taken into account, the distribution of lens masses derived from the ensemble of lensing events will be skewed toward lower values than the true underlying distribution. Other effects of blending are to significantly influence the rate of observable microlensing events by luminous stars, of binary stars, and of stars in crowded fields. In spite of the complications, blending may provide a way to learn more about galactic dynamics through the detailed study of events in which the additional light is contributed by a luminous star. Because the complications are ubiquitous and lead to significant effects, further study is required. Work to sharpen our tools for the identification of lensing events and to simulate the effects of blending on realistic observational programs is underway.

We would like to thank Emilio Falco, Shude Mao, Ramesh Narayan, and Paul Schechter for discussions.

# 4 References

Alcock, C., et al. 1993, Nature,365, 621
Alcock, C., et al. 1995 in preparation
Aubourg, E., et al. 1993, Nature, 365, 623
Bolatto, A. D., & Falco, E. E. 1994, ApJ, 436, 112
Griest, K., & Hu, W. 1992, ApJ, 397, 362
Kamionkowski, M. 1994, preprint
Miller, G. E., & Scalo, J. M. 1979 ApJS, 41, 513
Udalski, A., et al. 1993, Acta Astron., 43, 289
Udalski, A., Szymański, M., Mao, S., Di Stefano, R. I., Kałużny, J., Kubiak, M., Mateo, M., Krzemiński, W. 1994, ApJ, 436, L103